\documentclass[12pt]{article}
\usepackage{pic03}
\usepackage{hyperref}
\usepackage{url}
\usepackage{graphicx}

\begin{document}

\newcommand{\dzero}{\mbox{D\O} }
\newcommand{\pb}{pb$^{-1}$ }
\newcommand{\gevc}{GeV/c }
\newcommand{\gevcc}{GeV/c$^2$ }


\title{\bf SEARCHES FOR NEW PARTICLES AT THE ENERGY FRONTIER AT THE TEVATRON}
\author{Patrice VERDIER\\
{\em LAL, Universit\'e Paris-Sud, 91898 Orsay Cedex, France}}
\maketitle

%
%
%
%
%
%
\vspace{4.5cm}
%

\baselineskip=14.5pt
\begin{abstract}
Run~2 at the Tevatron started in spring 2001. CDF and \dzero are taking 
data at a center-of-mass energy of 1.96~TeV. First results on searches for 
phenomena beyond the Standard Model are presented. In January 2003, the 
integrated luminosity recorded per experiment was lower than the luminosity 
collected at Run~1. Nevertheless, these results are already
competitive due to improved detector capabilities and to 
the increase in the center-of-mass energy.
\end{abstract}
\newpage

\baselineskip=17pt

\section{Introduction}

The CDF and \dzero experiments are taking data at a center-of-mass energy of
1.96~TeV since the beginning of Run~2 at the Tevatron in March 2001.
130~\pb were delivered between January 2002 and January 2003. The integrated
luminosity used for these first results is between 30 and 80~\pb depending on 
the data sample. Both experiments are searching for new phenomena in a large
variety of channels. With a lower luminosity, Run~2 results starts to be
competitive with those of Run~1 thanks to higher center-of-mass energy and 
upgraded detector capabilities.

\section{Supersymmetry}

\subsection{SUGRA jets and missing $E_T$ search}

\begin{figure}[h!]
\begin{center}
\includegraphics[width=8cm,height=5.5cm]{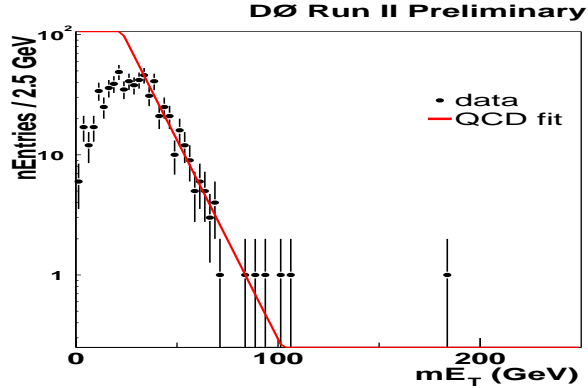}
\caption{\it Search for jets and missing $E_T$ at \dzero : missing $E_T$ 
distribution for data (dot) and the QCD fit (line).
\label{susyd0jet}}
\end{center}
\end{figure}

In SUGRA models with R-parity conservation, the Lightest Supersymmetric
Particle  (LSP) is heavy, neutral, stable and weakly interacting. The cascade
decays of  the squarks and gluinos into quarks, gluons and the LSP lead
therefore to the  signature of jets and missing transverse energy (missing
$E_T$). At the Tevatron, this signal  has the highest cross section. However, the
main background is instrumental and  comes from the tail of the Standard
Model jet production (QCD), which largely dominates. A preliminary analysis
using 4~\pb of \dzero data shows that  the backgrounds can be controlled and
understood. A high $p_T$ jet above 65~\gevc is required at the trigger level.
This cut is reinforced offline to 100~\gevc where the trigger efficiency reaches
100\%. The QCD background is fitted at low missing $E_T$ and extrapolated to
higher values. Figure~\ref{susyd0jet} shows the missing $E_T$ distribution
and the agreement with the QCD fit. For a missing $E_T$ cut at  100~GeV, the
number of selected events is 3 while the Standard Model expectation is 
2.7$\pm$1.8.

\subsection{SUGRA di- and trilepton search}

\begin{figure}[b!]
\begin{center}
\begin{tabular} {cc}
\includegraphics[height=5.5cm]{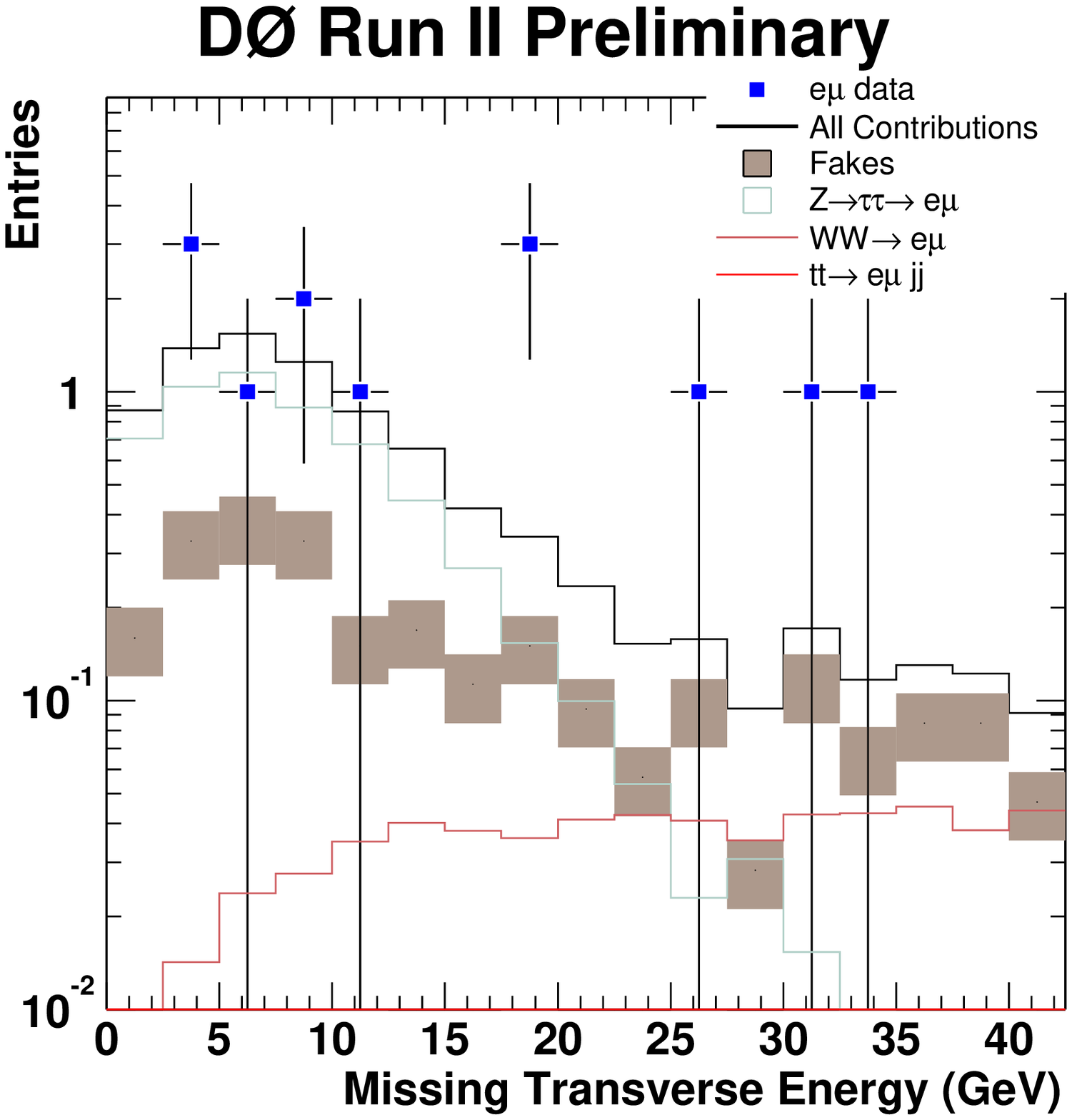} &
\includegraphics[height=5.5cm,width=6cm]{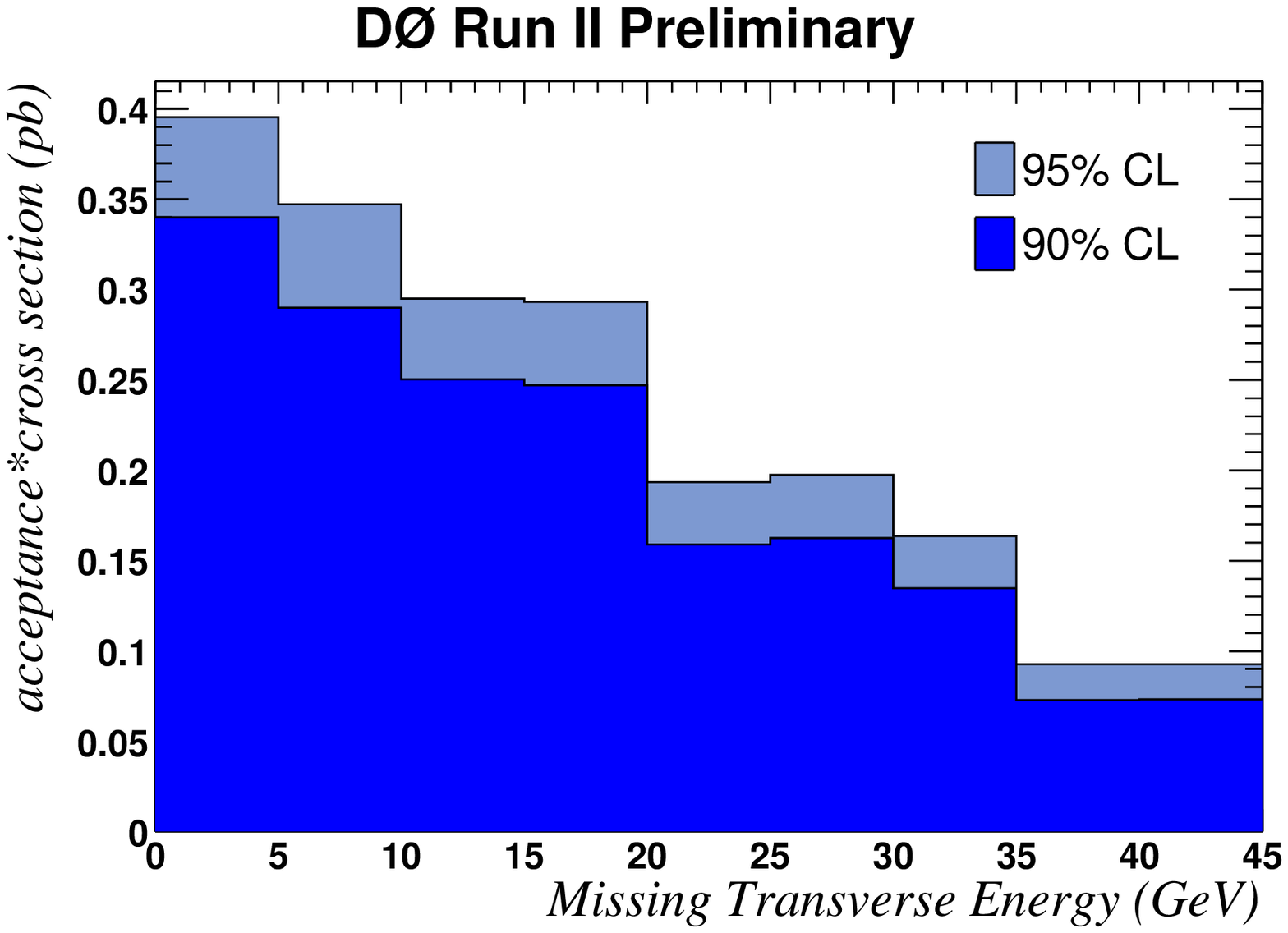} \\
\end{tabular}
\caption{\it Search for $e\mu$ at \dzero : Missing $E_T$ distribution (left) and
cross section limit as a function of the missing $E_T$ cut (right).
\label{exod0emu}}
\end{center}
\end{figure}

\begin{figure}[t!]
\begin{center}
\includegraphics[width=10cm,height=5.5cm]{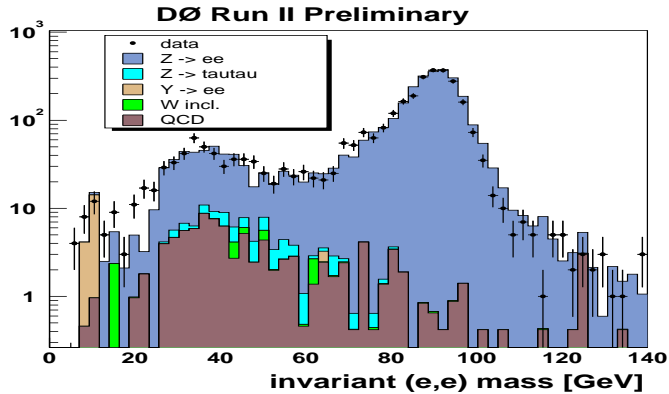}
\caption{\it Search for $eel$ at \dzero : dielectron invariant mass distribution.
\label{susyd0lep}}
\end{center}
\end{figure}

Gauginos are predicted to be light in SUGRA models and they have a clear 
multiple leptons signature at the Tevatron. However, this so-called golden
channel suffers from a much lower cross section than squarks and gluinos
ones.\\ On 30~\pb of data, \dzero is performing a model independent analysis in
the  channel with one electron and one muon. This signal has a very low
background coming from $WW$ and $t\bar{t}$ at high missing $E_T$ and 
$Z^0/\gamma^*\to\tau^+\tau^-$ at low missing $E_T$. One electron and one muon
both with $p_T > 15~\gevc$ are required.  The fake electron and muon fake rate
is estimated from the data and physical  backgrounds are estimated with
simulation. The missing $E_T$ distribution in Figure~\ref{exod0emu}
shows a good agreement between data and Standard Model expectation. 
A model independent upper limit on acceptance times cross section is set for 
new signal leading to $e\mu$ fnal states a function of the missing $E_T$ 
cut.\\
In the channel with two electrons and a third lepton, 40~\pb of \dzero data 
are analyzed. Selected events must contain at least two electrons with  $p_T$
greater than 15~\gevc for the first one and greater than 10~\gevc for the
second one. Figure~\ref{susyd0lep} shows the di-electron invariant mass at
this level of the analysis. The search is restricted to di-electron invariant
mass between 10 and 70~\gevcc and to events with transverse mass greater than
15~\gevcc. Finally, a third isolated track is required in the pseudo-rapidity 
range $|\eta| < 3$ before requiring the missing $E_T$ to be greater than 15~GeV.
No events are selected while the background is estimated to be 0.0$\pm$1.4. The
selection efficiency for a typical SUGRA signal at the edge of the current
exclusion limit is 2-4\%. It is not sufficient to extend the excluded region
in the SUGRA parameter space. Higher luminosity is needed for this channel
while the analysis is being improved.

\subsection{GMSB photon search}

\begin{figure}[ht!]
\begin{center}
\begin{tabular} {cc}
\includegraphics[height=5.5cm]{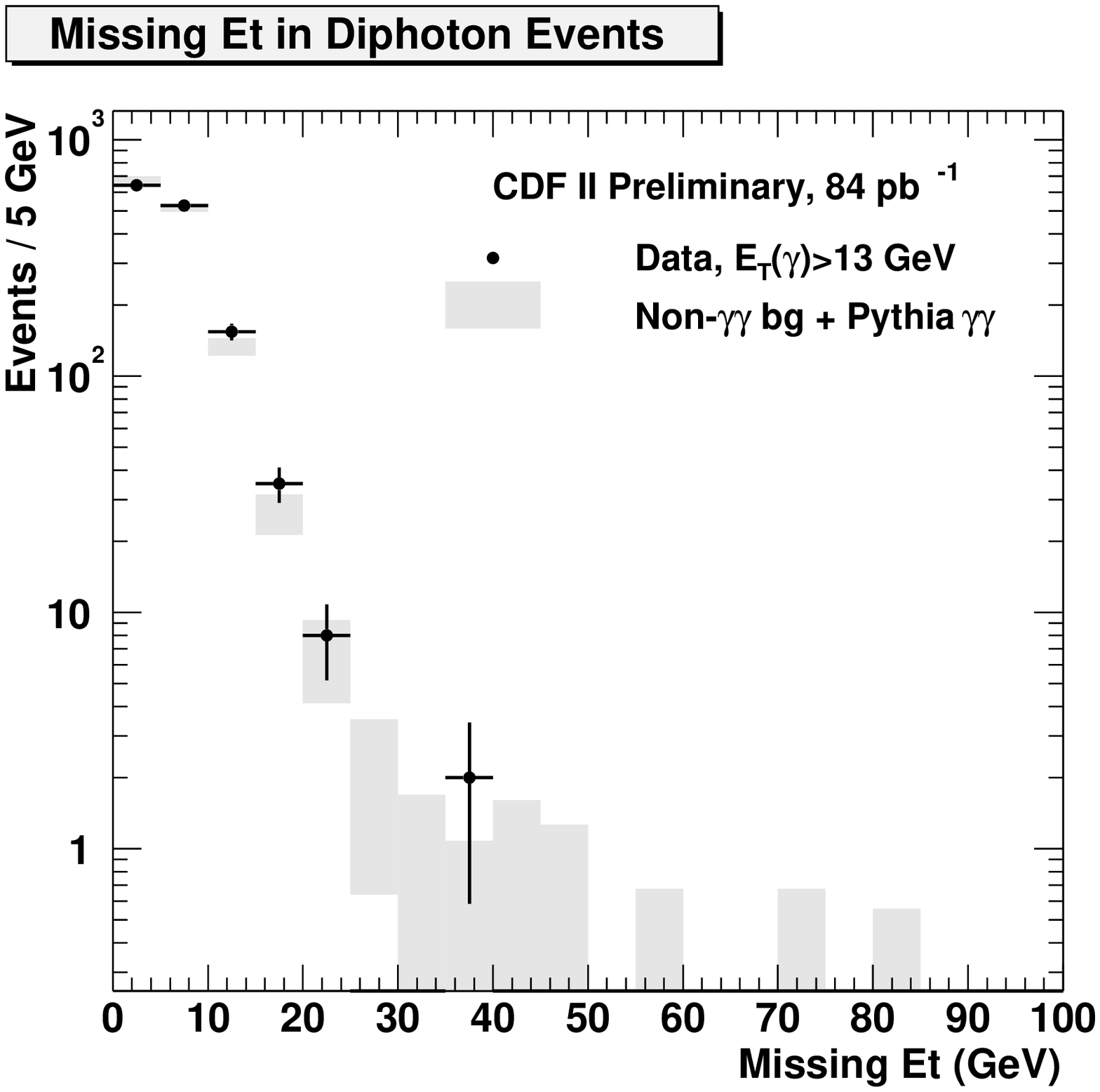} &
\includegraphics[height=5.5cm]{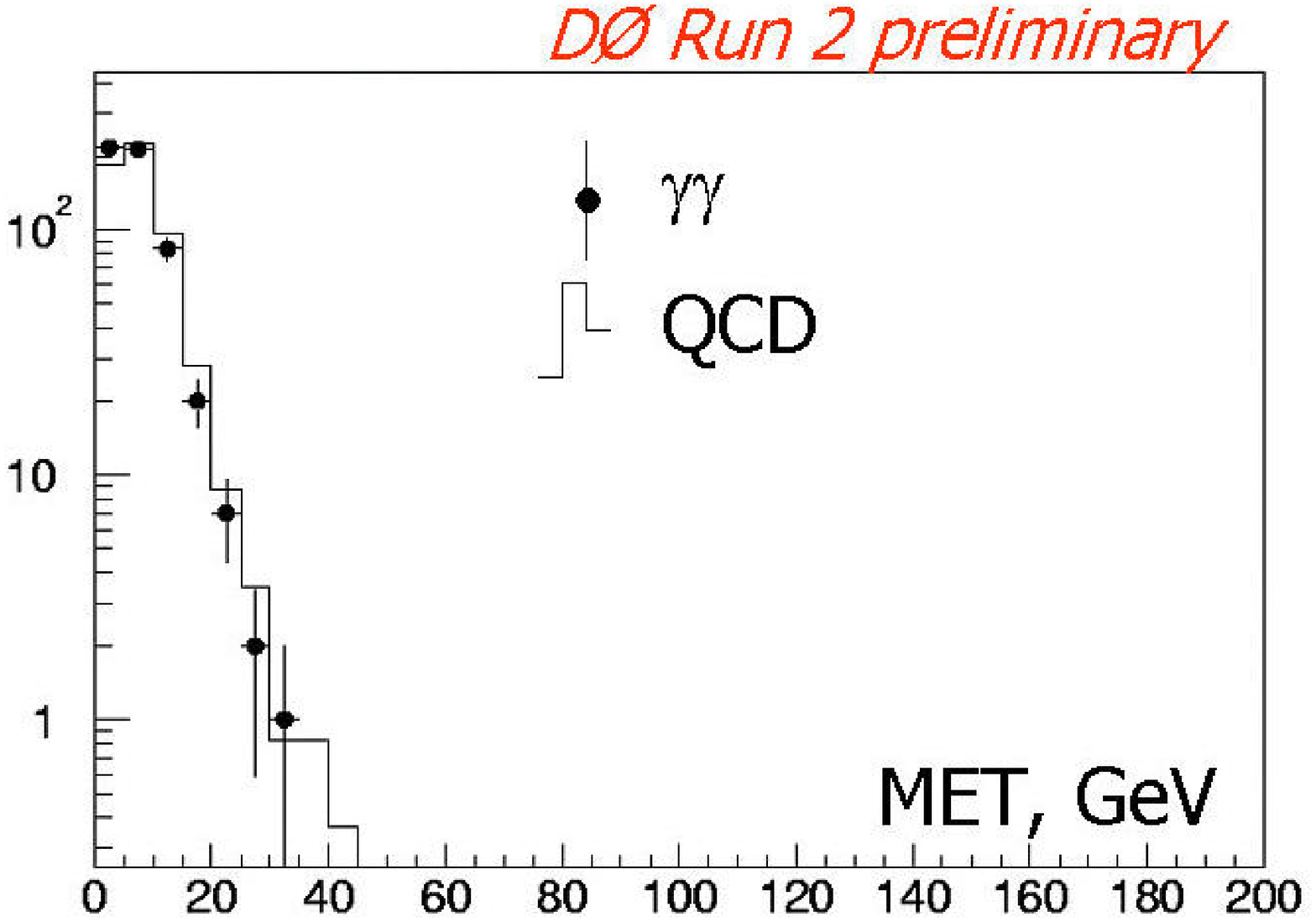} \\
\end{tabular}
\caption{\it Search for photon events in GMSB at CDF and \dzero : Missing 
$E_T$ distribution for data (points) and backgrounds (histogram) in the CDF 
(left) and \dzero (right) analyses.
\label{susyphot} }
\end{center}
\end{figure}

In Gauge Mediated Supersymmetry Breaking (GMSB), the LSP is a light gravitino
with a mass which could be much lower than the $eV$. If the next-to-lightest
supersymmetric particle is a bino-like neutralino, it would decay into a
gravitino and a photon. GMSB signature at the Tevatron is therefore two photons
and missing $E_T$.\\ Requiring two central photons with $E_T > 13 GeV$, CDF
selects 1365 events from 81~\pb of data. \dzero is using a data sample of
50~\pb. Events must have two photons with $E_T > 20 GeV$. The main background
comes from QCD fake events and is determined from data. Figure~\ref{susyphot}
shows the missing $E_T$  distribution of these events and the good agreement
with the background  expectation. The limits at 95\% confidence level on the
GMSB scale $\Lambda$  using the Snowmass slope~\cite{snowmass} ($M = 2\Lambda ,
N_5 = 1, tan  \beta = 15$ and $\mu > 0$) are 51~TeV for \dzero and 50~TeV for
CDF. These results are very close to Run~1 results.

\section{Leptoquarks}

At the Tevatron, leptoquarks would be pair produced through gluon fusion or 
quark anti-quark annihilation. Due to experimental constraints, they are
expected to couple only to fermions of the same generation. 

\subsection{First generation leptoquarks}

\begin{figure}[t!]
\begin{center}
\begin{tabular} {cc}
\includegraphics[height=5cm]{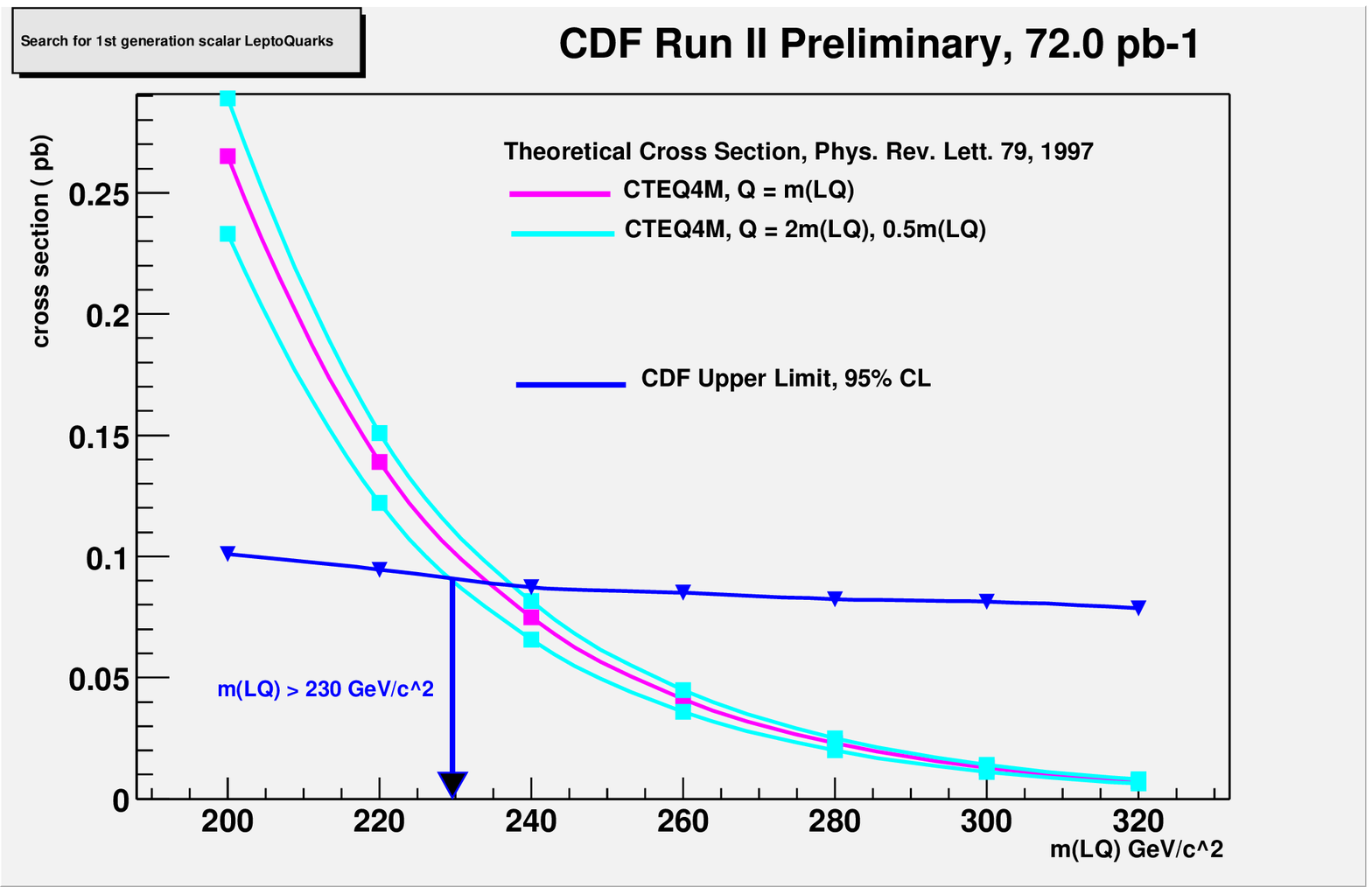} &
\includegraphics[height=5cm]{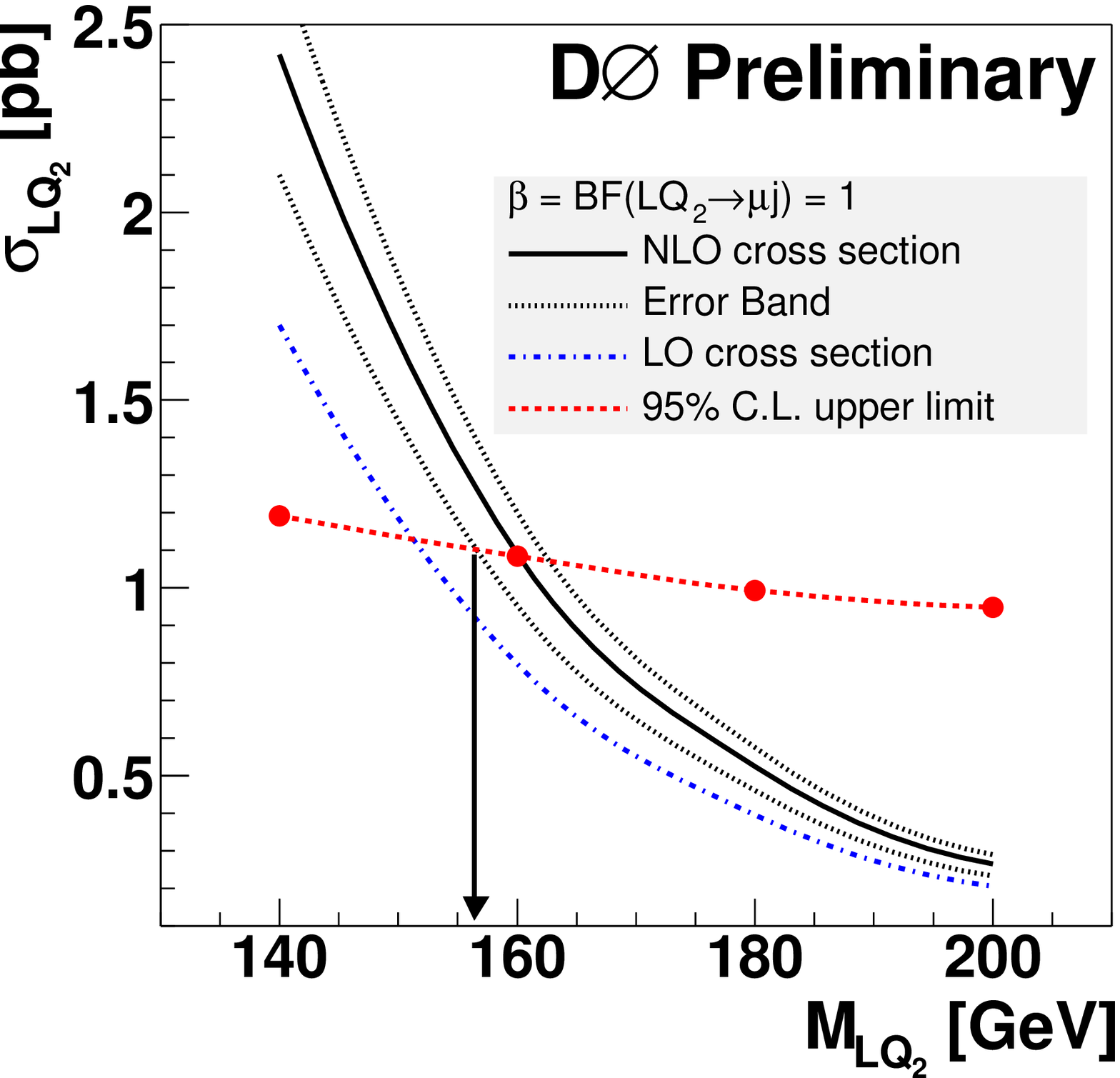} \\
\end{tabular}
\caption{\it 
\label{leptoquark} Search for leptoquarks : 95\% cross section limit as a
function of the leptoquark mass for the CDF analysis $LQ_1LQ_1\to eeqq$ (left)
and for the \dzero analysis $LQ_2LQ_2\to \mu\mu qq$ (right).}
\end{center}
\end{figure}

CDF investigated the following decay of leptoquarks : $LQ_1LQ_1\to eeqq$.
Events are selected if they contains two electrons with $E_T >$~25~GeV, and
two jets with $E_T >$~30~GeV for the first one, and $E_T >$~15~GeV for the
second one. Then topological cuts are applied and no events from 72~\pb of 
data survive the analysis cuts. The Standard Model background is expected to 
be 3.4$\pm$3.0.\\
\dzero also searched for leptoquarks in this decay channel. The data sample
corresponds to 43~\pb. The analysis is very similar to the one of CDF. No
events are selected for an expected background of 0.08$\pm$0.02.\\
95\% confidence level (CL) limits on the first generation leptoquark cross
section  are derived (Figure~\ref{leptoquark}). These can be translated into
a lower limit on the leptoquark mass :  $M(LQ_1)>$ 230~\gevcc for CDF, and 
$M(LQ_1) > $~179~\gevcc for \dzero. These results already improved those of
Run~1.

\subsection{Second generation leptoquarks}

Another topology comes from second generation leptoquark decay :
$LQ_2LQ_2\to \mu\mu qq$. In 30~\pb of \dzero data, events must contain two
opposite sign muons with $p_T > 15$~\gevc, two jets with $p_T > 20$~\gevc,
and the invariant dimuon mass is required to be greater than 110~\gevcc.
The dominant background comes from $Z^0/\gamma^* \rightarrow \mu \mu$
accompanied by jets due to radiation. No events are found, and a cross section
limit is derived (Figure~\ref{leptoquark}). The lower limit on the leptoquark
mass is 157~\gevcc. It is still far from the Run~1 limit of 200~\gevcc.

\section{Exotics}

\subsection{Charged massive stable particles}

\begin{figure}[t!]
\begin{center}
\begin{tabular} {cc}
\includegraphics[width=7cm]{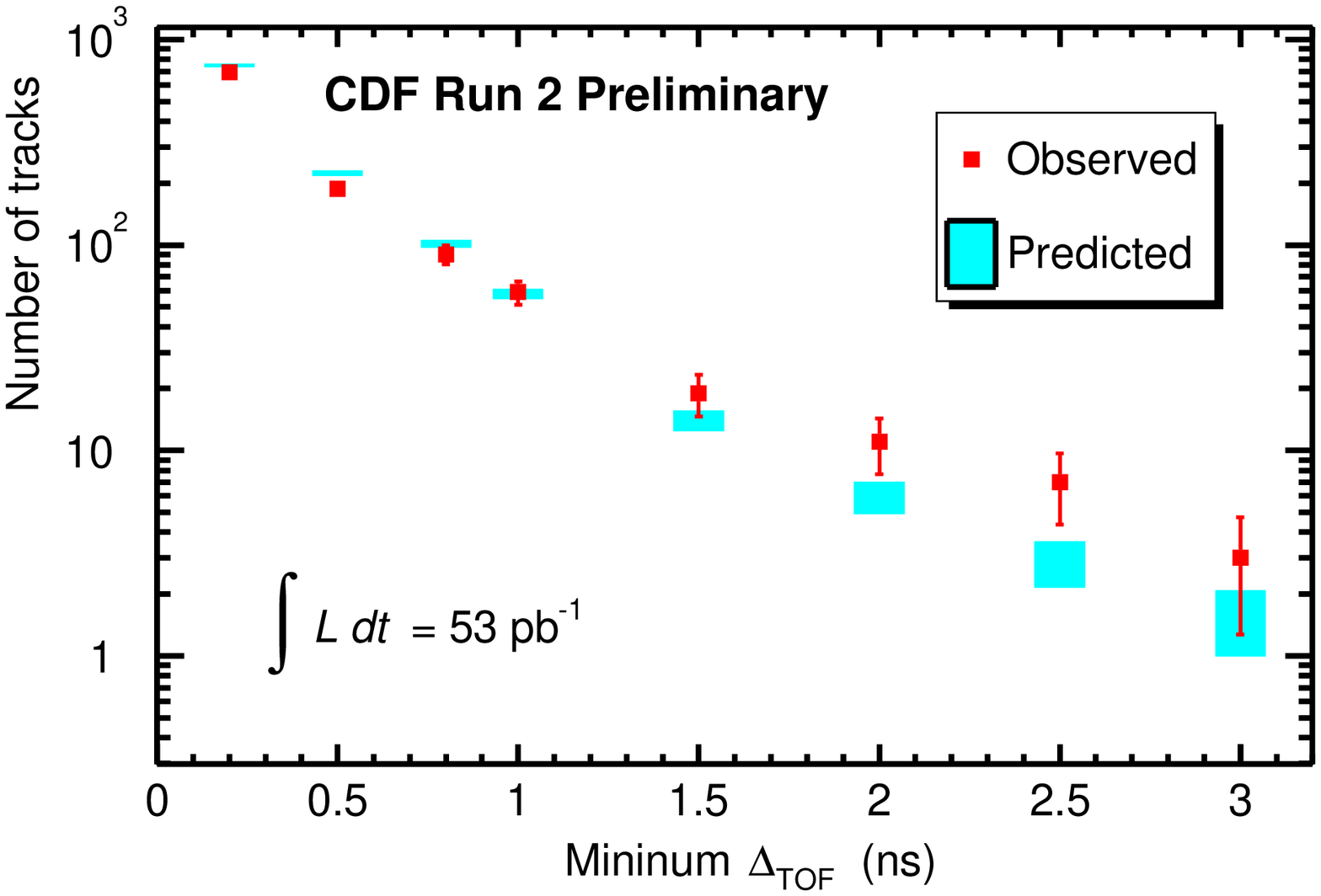} &
\includegraphics[width=7cm]{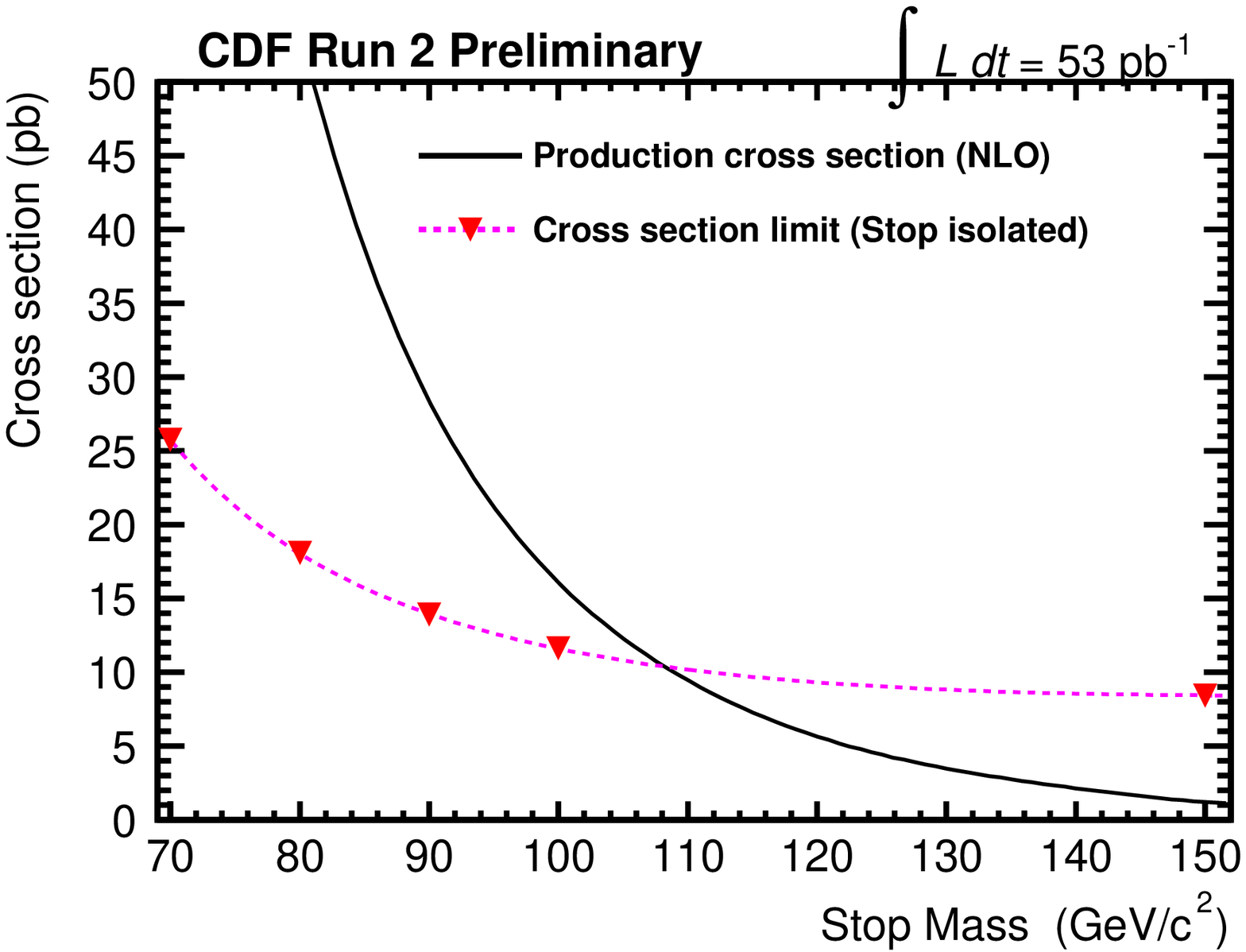} \\
\end{tabular}
\caption{\it Search for charged massive stable particles at CDF : $\Delta_t$
distribution measured with the time of flight detector (left), cross section
upper limit as a function of the stop mass (right).
\label{exocdfstable} }
\end{center}
\end{figure}

The new Time of flight system of CDF provides a better sensitivity to 
$\beta\gamma$ than a $dE/dx$ measurement. It is therefore used to search for
charged massive particles long lived enough to  escape the detector. A data
sample of 52~\pb coming from high $p_T$ muon trigger is used, and an  offline
$P_T$ cut at 40~\gevc is  applied in order to have full tracking efficiency.
The time $t_0$ at which the interaction occurred is obtained from tracks with
$P_T < 20$~\gevc. Tracks with high $\Delta_t = t_{tracks} - t_0$ are searched
for.  Figure~\ref{exocdfstable} shows the distribution of $\Delta_t$. 
Optimized to maximize the discovery probability, the final cut at 2.5~ns keeps
7 events. The background is estimated using tracks with  $20 < P_T < 40$~\gevc
: $2.9\pm 0.7(stat.)\pm 3.1 (syst.)$.\\ In the stable stop scenario,  cross
section upper limits are derived (Figure~\ref{exocdfstable}). The resulting
mass limit is $M(stop) >$~ 108~\gevcc.

\subsection{Search for resonances in dijets}

\begin{figure}[htb]
\begin{center}
\includegraphics[width=6cm]{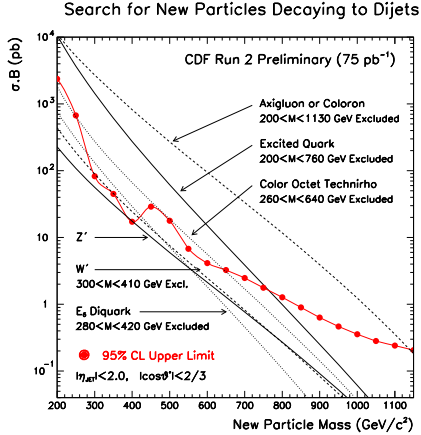}
\caption{\it Search for resonances in dijets at CDF : 95\% CL upper limits on
the cross section times branching ratio for new particles decaying into
dijets. 
\label{exocdfdijet} }
\end{center}
\end{figure}

A search for new particles decaying to dijets is performed with 75~\pb of CDF
data. The dijet invariant mass spectrum is obtained with the two highest $E_T$
jets of each event. It is fitted with a simple background parametrization and 
no significant evidence for a signal is observed. 95\% CL upper limits on the
cross section times branching ratio are set and compared with the prediction
for axigluons,  flavor universal colorons, excited quarks, Color Octet \
Technirhos, E6 diquarks, W' and Z' (Figure~\ref{exocdfdijet}).
CDF first Run~2 results improved Run~1 limits and allow to exclude :
\begin{itemize} 
\item{} axigluons or flavor universal colorons
for masses between 200 and 1130 GeV.
\item{}excited quarks with mass between
200 and 760 GeV. 
\item{} color octet technirhos between 260 and 640 GeV.
\item{} E6 diquarks with mass between 280 and 420 GeV. 
\item{} W' with mass between 300 and 410 GeV. 
\end{itemize}

\subsection{Extra Gauge Bosons}
\label{refzprime}

\begin{figure}[p!]
\begin{center}
\begin{tabular}{cc}
\includegraphics[height=6cm,width=7cm]{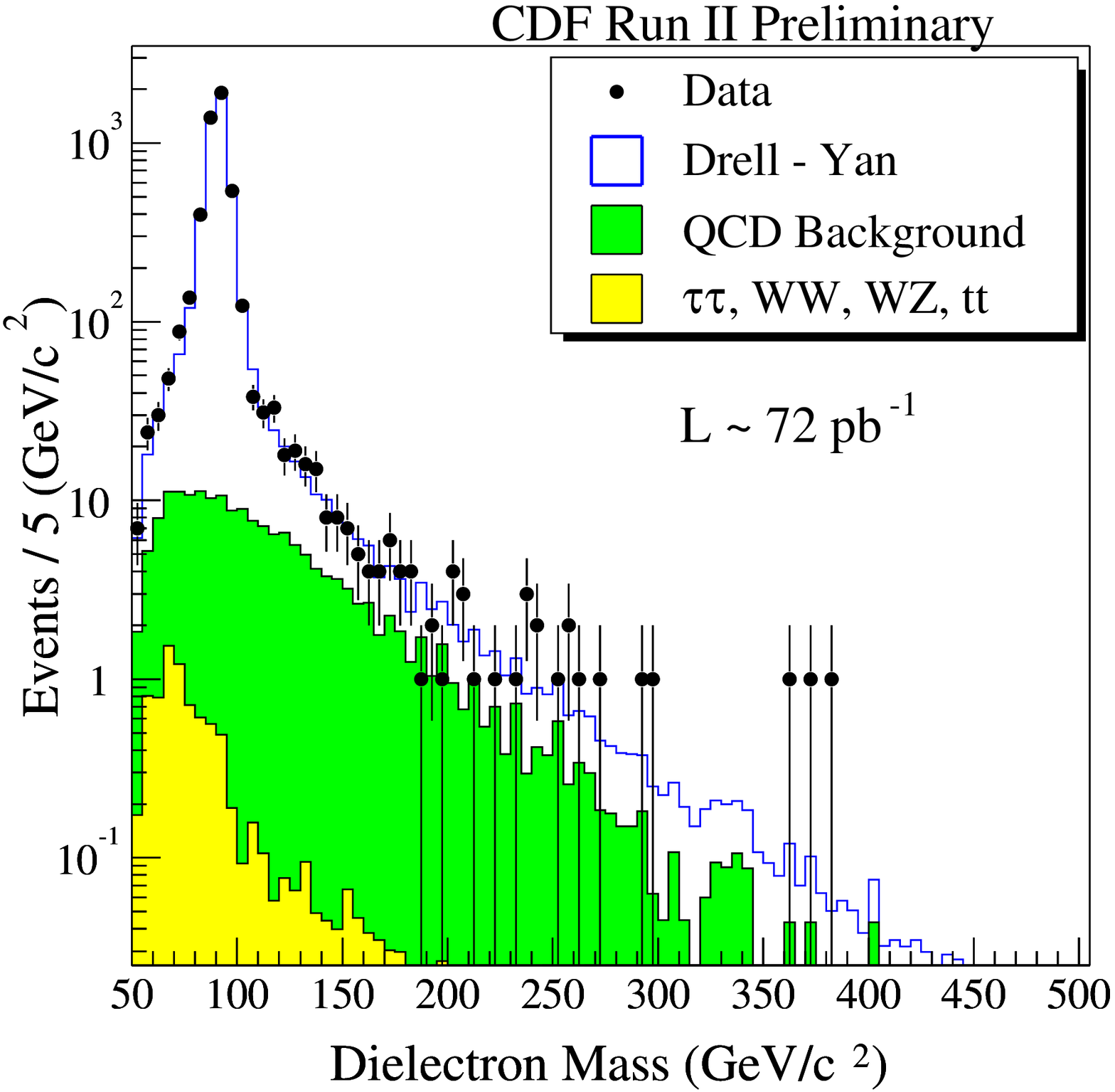} &
\includegraphics[height=6cm,width=7cm]{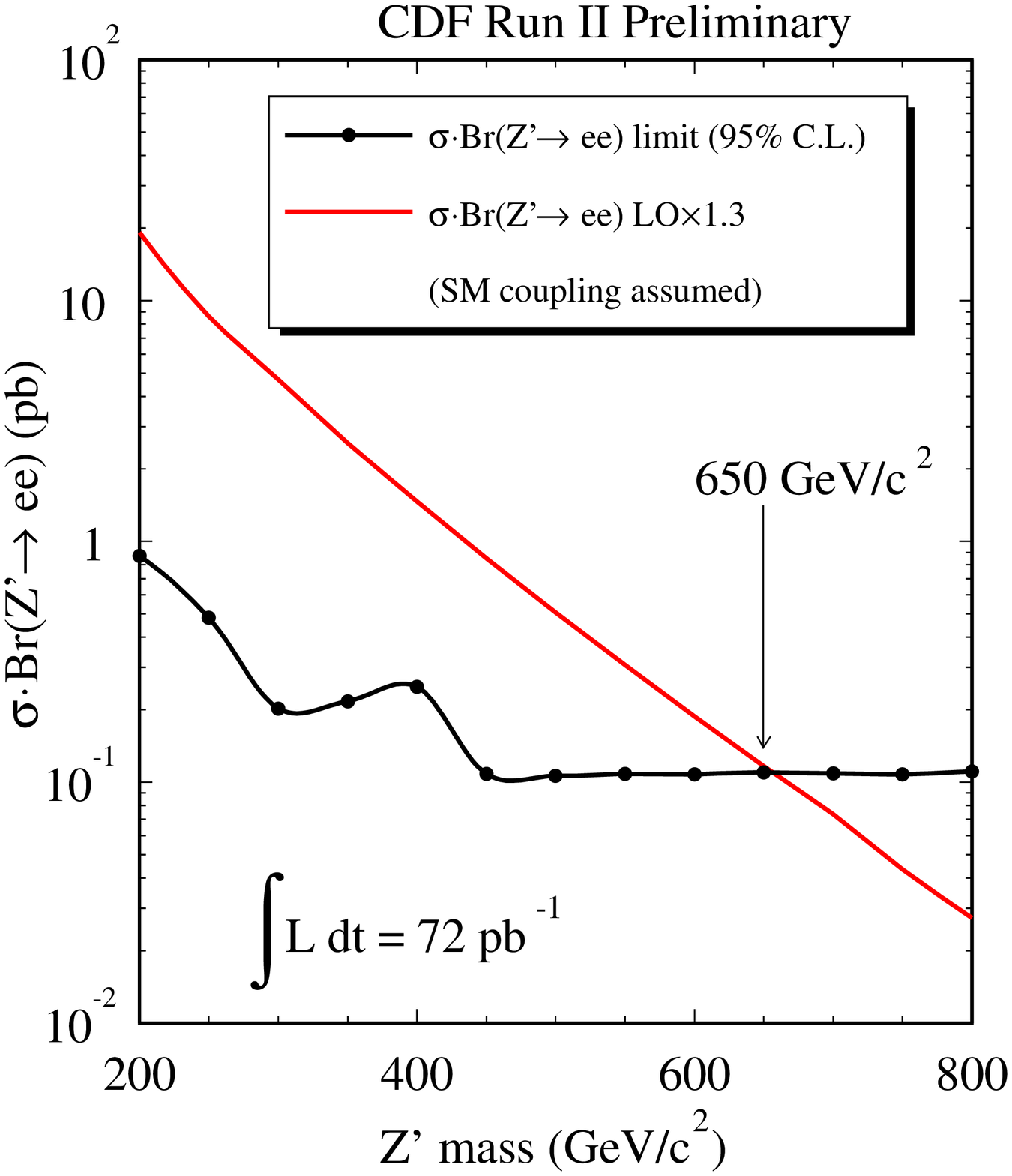} \\
\includegraphics[height=6cm,width=7cm]{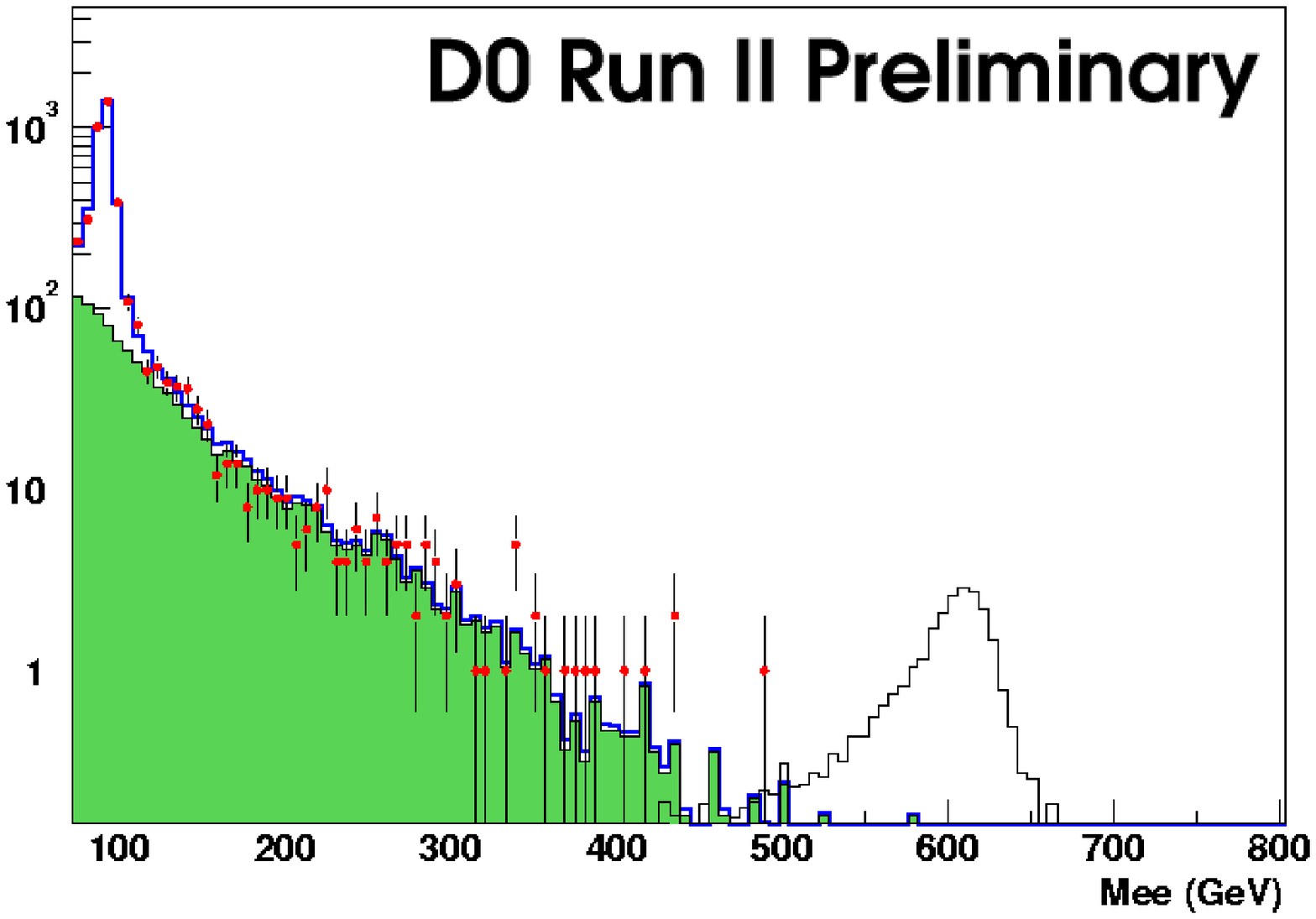} &
\includegraphics[height=6cm,width=7cm]{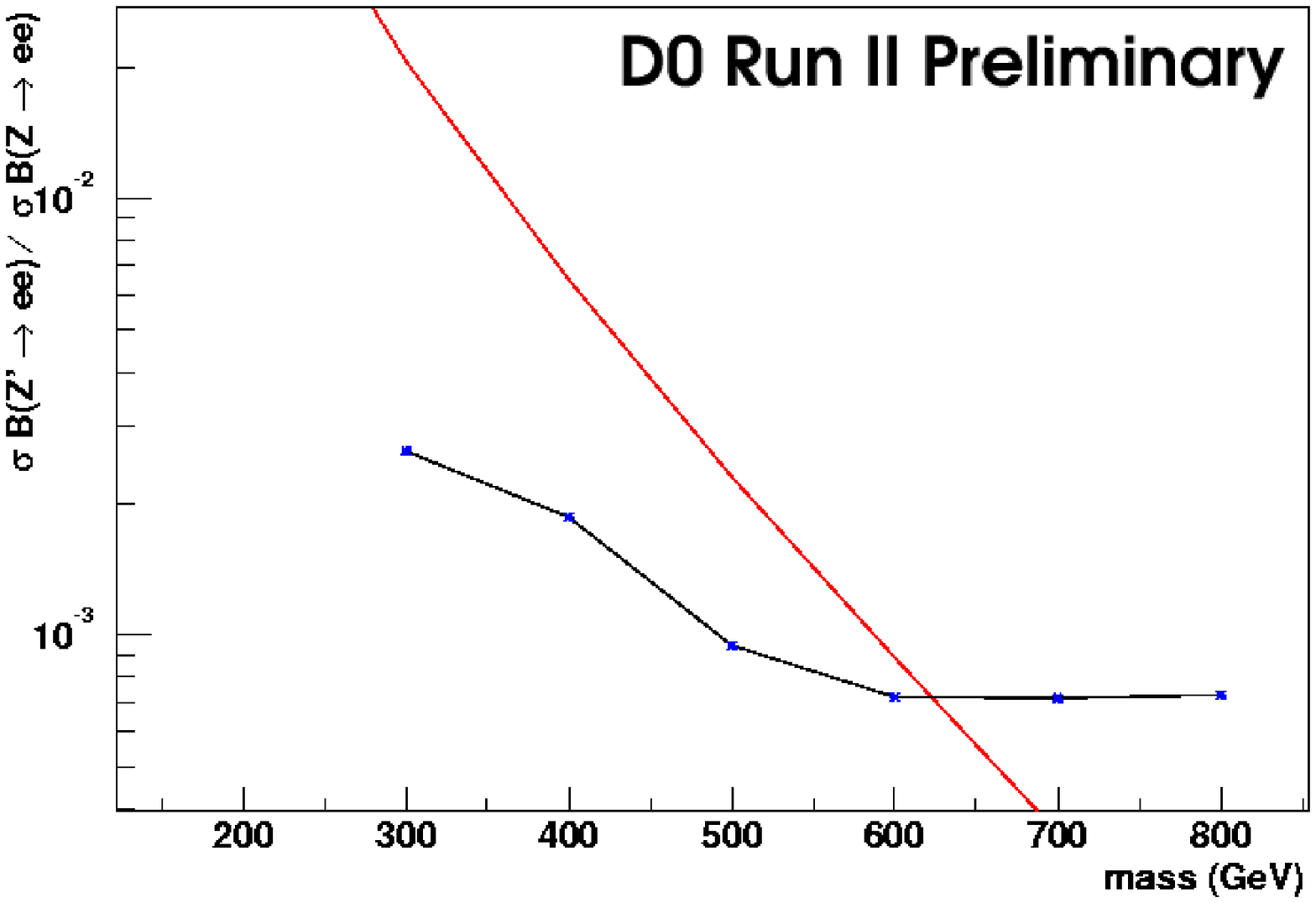} \\
\includegraphics[height=6cm,width=7cm]{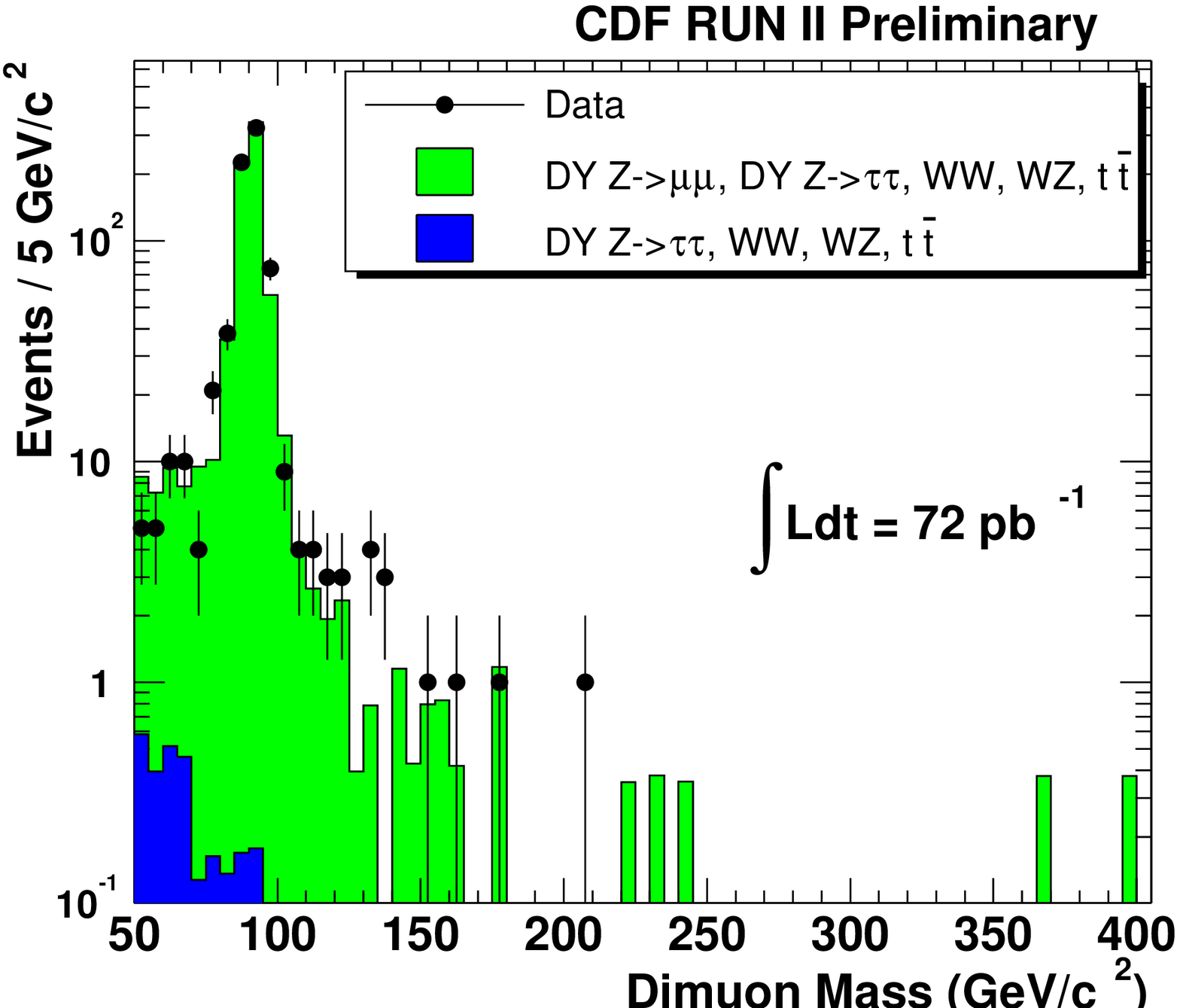} &
\includegraphics[height=6cm,width=7cm]{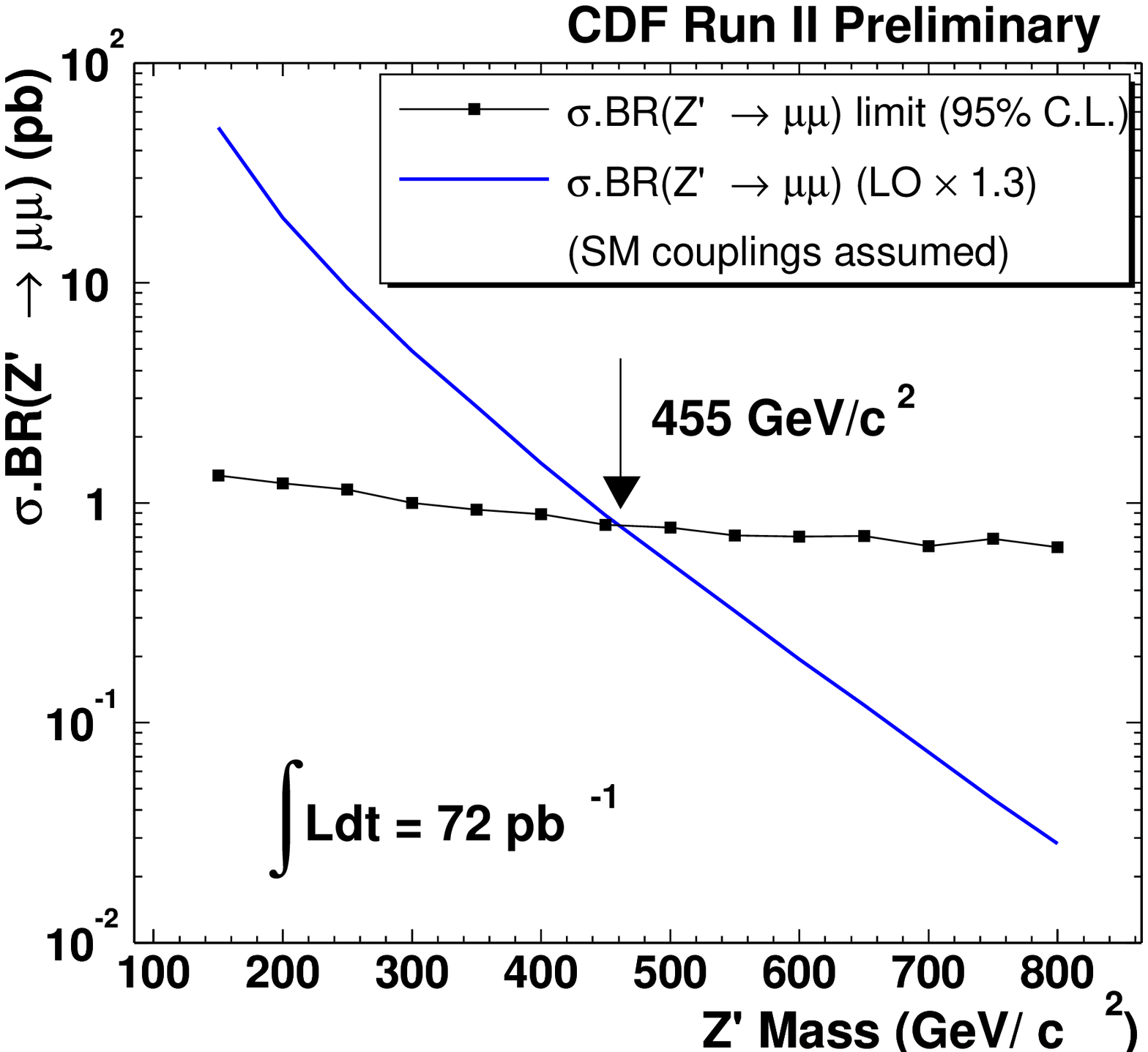} \\
\end{tabular}
\caption{\it Search for extra gauge bosons $Z'$ : Drell-Yan dilepton invariant 
mass spectra (left) and the corresponding cross section limit (right) of
CDF dielectron analysis (upper), of \dzero dielectron analysis (middle) and of
CDF dimuon analysis (lower).
\label{exozprime}}
\end{center}
\end{figure}

CDF and \dzero are searching for a new neutral gauge boson $Z'$ using the
Drell-Yan dilepton mass spectrum. Such high mass particles would be produced  by
quark-antiquark annihilation and would decay into a pair of opposite sign
leptons.\\ 
For the decay channel $Z'\to e^+e^-$, CDF analysis requires two good
electrons with $P_T$ greater than 25~\gevc either in the central or in the plug
calorimeter and additional cuts are imposed to remove the remaining  $W+jets$
background. \dzero selects events with two good electrons with  $P_T$ greater
than 25~\gevc but restricts the selection to the pseudo-rapidity range
$|\eta|<1.1$. Figure~\ref{exozprime} shows  the obtained dielectron mass
distribution and the good agreement between data and Standard Model
expectation.\\ 
CDF also performed the search for $Z'$ boson in the muon
channel  $Z'\to \mu^+\mu^-$. It is based on the selection of two good muons
with  $P_T>20$~\gevc. Cosmic and QCD background are removed with cuts on the
track impact parameter and on the muon isolation. The dimuon mass spectrum is
shown in Figure~\ref{exozprime}.\\ 
Assuming Standard Model couplings, 95\% CL
lower limits are set on the $Z'$  boson mass : 650~\gevcc by CDF and 620~\gevcc
by \dzero in the electron channel, 455~\gevcc by CDF in the muon channel.

\section{Extra Dimensions}

\subsection{\it Search for small extra dimensions}

\begin{figure}[htb]
\begin{center}
\begin{tabular} {cc}
\includegraphics[height=6cm]{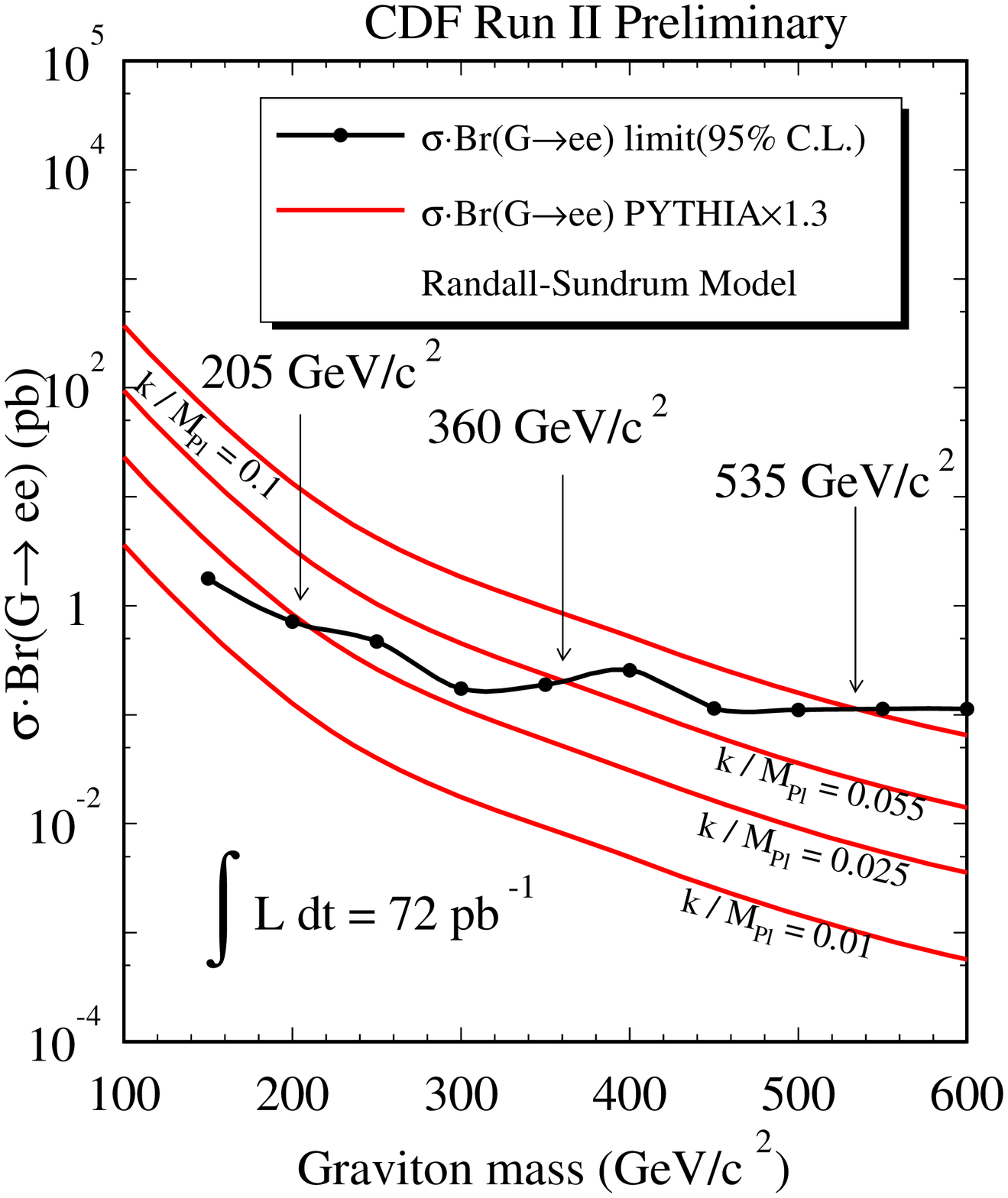} &
\includegraphics[height=6cm]{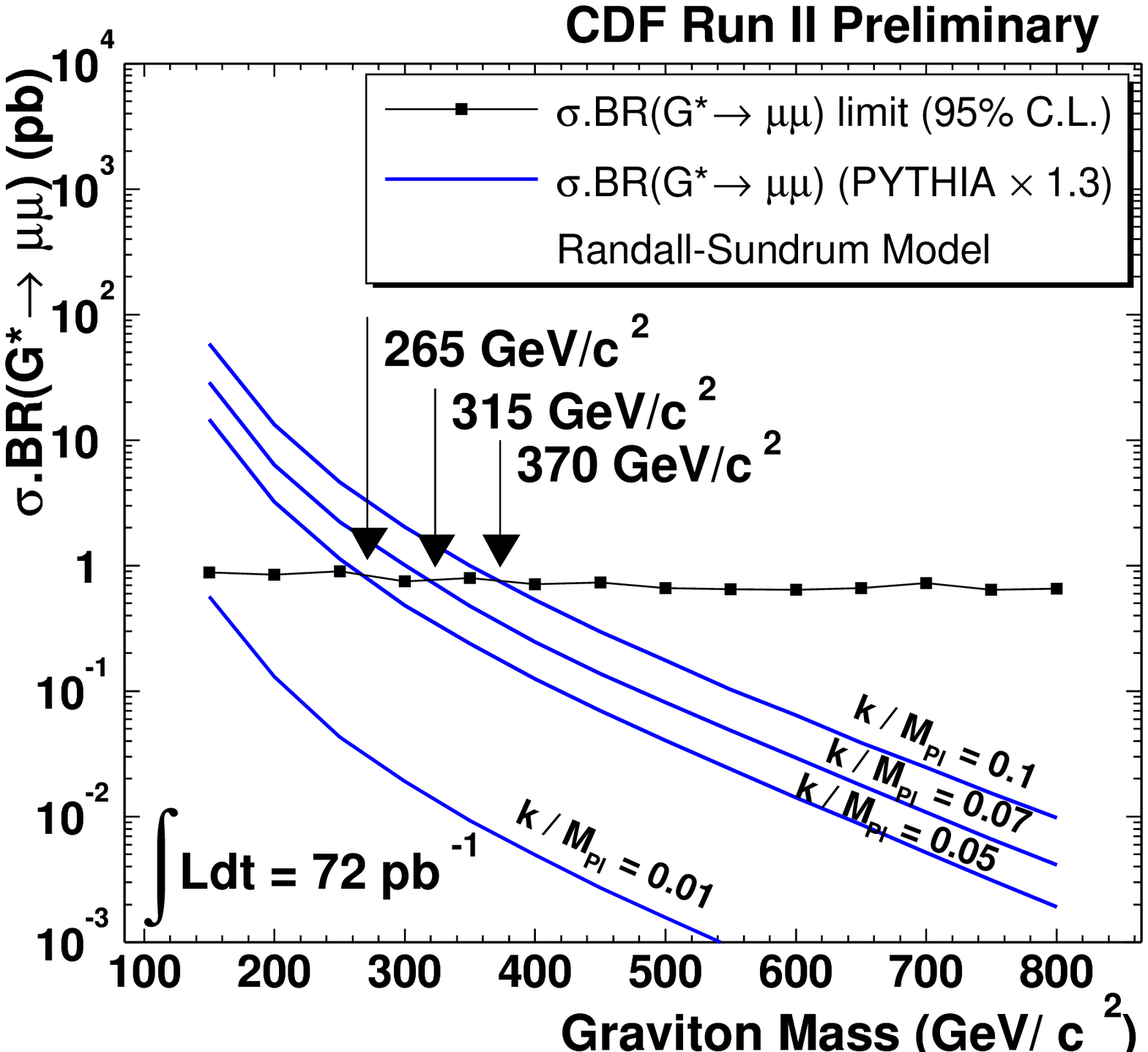} \\
\end{tabular}
\caption{\it Search for small extra dimensions at CDF : Cross section limits as
a function of the Randall Sundrum graviton mass in the dielectron (left) and
dimuon (right) channels.
\label{edcdflep}}
\end{center}
\end{figure}

Extra dimensions have been recently introduced to solve the hierarchy
problem\cite{add}. In the Randall-Sundrum graviton model\cite{rs}, the Kaluza 
Klein excitations of the graviton can be separately produced as resonances, 
enhancing the Drell-Yan cross section at large mass. CDF interprets the high 
dilepton mass analysis~\ref{refzprime} in this model. The results are lower
mass limits on the graviton mass as a function of the graviton mass coupling
$k/M_{PL}$ (Figure~\ref{edcdflep}).

\subsection{\it Large Extra Dimensions}

\begin{figure}[h!]
\begin{center}
\includegraphics[width=10cm]{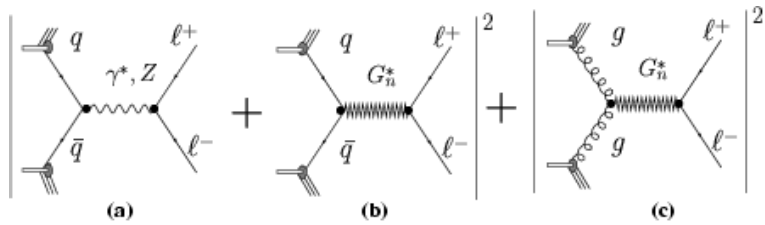}
\caption{\it 
\label{edd0feyn} Feynman diagrams showing contributions of virtual graviton
exchange to Drell-Yan processes.}
\end{center}
\end{figure}

The ADD model~\cite{add} predicts an excess of 
high-mass dielectron, diphoton or dimuon events due to the coupling to 
Kaluza-Klein gravitons. Feynman diagrams of the Standard Model and graviton 
contributions to dilepton final states are shown in Figure~\ref{edd0feyn}. The
differential cross section can be parametrized as:
\begin{equation}
\frac{d^2 \sigma}{dM d\cos\theta^*} = f_{SM} + f_{interf} \eta_G +
                                     f_{KK} \eta_G^2 .
\end{equation}
where $\cos\theta^*$ is the scattering angle in the di-em or dimuon rest frame,
and $\eta_G$ measures the contribution from gravitons.\\

\begin{figure}[h!]
\begin{center}
\includegraphics[width=9cm]{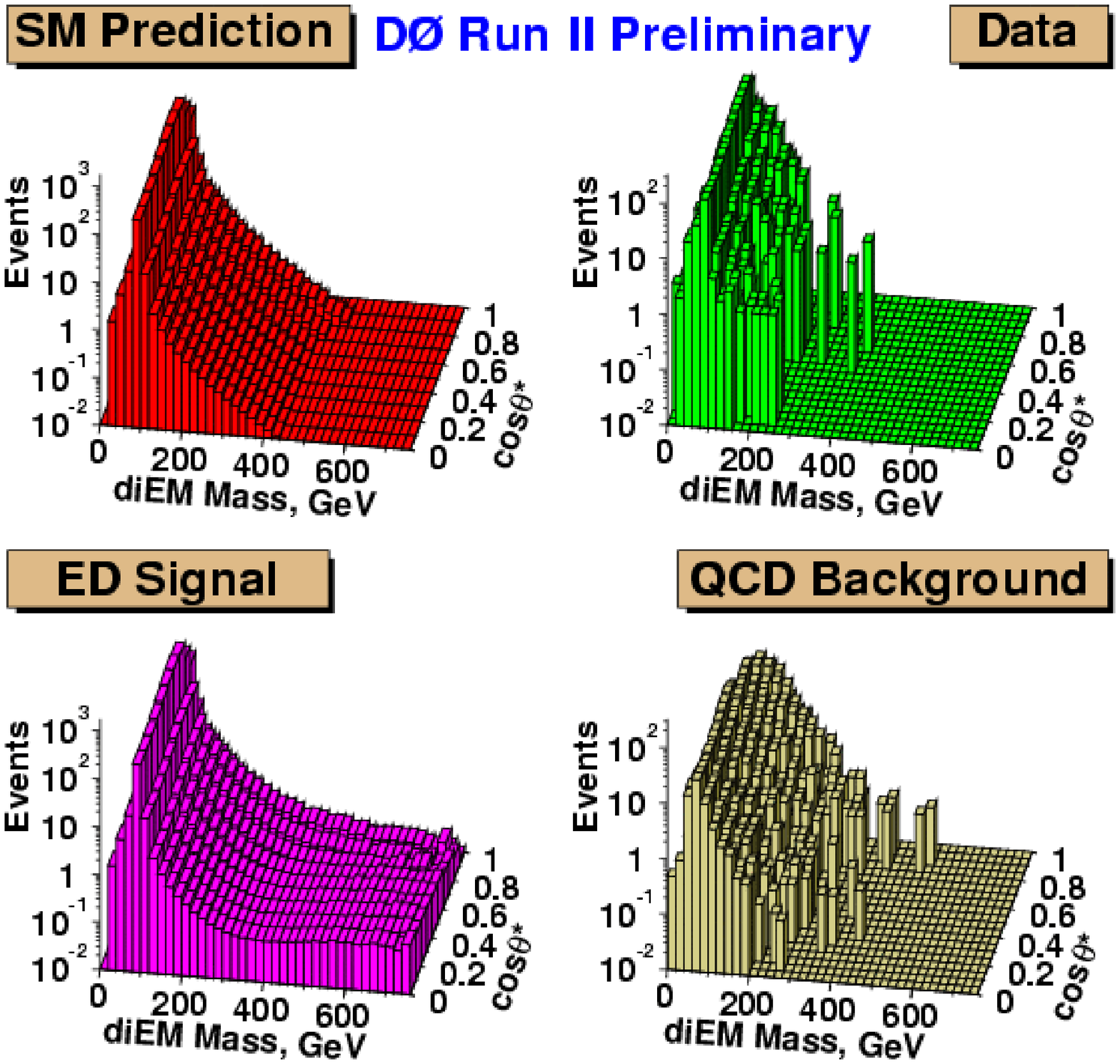}
\caption{\it Large Extra Dimension search at \dzero in the diEM channel :
invariant mass distribution as a function of $\cos\theta^*$ for Standard Model
expectation, data, LED signal and QCD background.
\label{edd0em}}
\end{center}
\end{figure}

\begin{figure}[htb]
\begin{center}
\includegraphics[width=10cm]{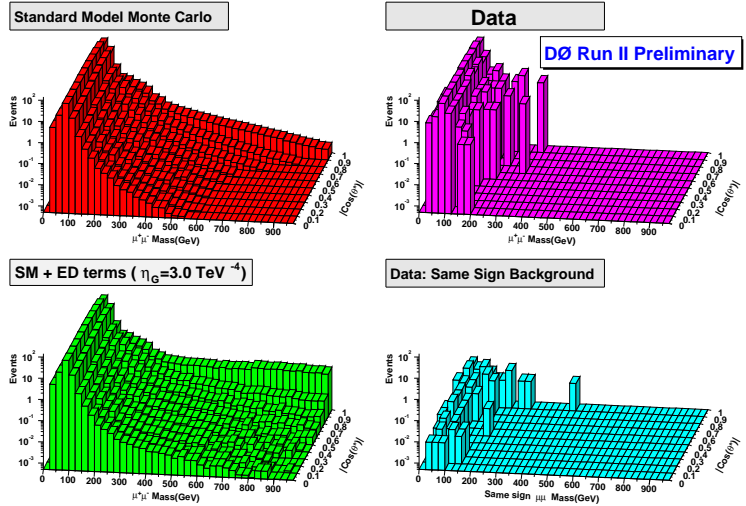}
\caption{\it Large Extra Dimension search at \dzero in the dimuon channel : 
invariant mass distribution as a function of $\cos\theta^*$ for Standard Model
expectation, data, LED signal and same-sign dimuon background.
\label{edd0mu} }
\end{center}
\end{figure}

\noindent
D\O\ is searching for LED. In the diEM analysis, no track requirement is 
imposed, and the dielectron and diphoton  channels are treated simultaneously. 
Events with 2 electromagnetic objects with $E_T >$25~GeV are selected. The
missing $E_T$ is required to be smaller than 25~GeV. In the dimuon channel,
events must have two isolated muons with $p_T >$15~\gevc matched with a central
track. The dimuon mass must be greater than 40~\gevcc. In both analyses, 
physics backgrounds are derived from simulation and instrumental backgrounds 
from data. The data distribution (Figures~\ref{edd0em} and~\ref{edd0mu}) are
fitted in  the invariant mass-$\cos\theta^*$ plane to the signal plus
background distribution. This procedure allows to set a lower limit on $\eta_G$
which is translated into a lower limit on $M_S$, the fundamental Planck scale.
Results are shown in Table~\ref{tab:led} for different LED formalisms.\\
The diEM limits are very close to the Run~1 results and the dimuon analysis is
new at the Tevatron.

\begin{table}
\caption{Large Extra Dimension search at \dzero : Lower limits in TeV on $M_S$,
the fundamental Planck scale, for various LED formalisms.
\label{tab:led}}
\vspace{0.4cm}
\centerline{
\begin{tabular}{|l|cccc|}
\hline
Formalism & GRW & HLZ, n=2 & HLZ, n=7 & Hewett, $\lambda = +1$ \\
\hline
\hline
Di-Em ($\approx 50 pb^{-1}$) & 1.12 & 1.16 & 0.89 & 1 \\
\hline
Dimuon ($\approx 30 pb^{-1}$) & 0.79 & 0.68 & 0.63 & 0.71 \\
\hline
\end{tabular}}
\end{table}

\section{Conclusion}

CDF and \dzero are searching for many signatures of physics beyond the
Standard Model in Run~2 data. These first results show the effects of improved 
detector capabilities and higher center-of-mass energy. Discovery
potential at the Tevatron will increase with higher luminosity, improved
analyses, better understanding of the detector and extended trigger 
capabilities.


\end{document}